\documentclass[12pt]{article}
\usepackage[dvips]{graphicx}
\usepackage{pdproc}
\usepackage{amsmath}
  \makeatletter 
  \def\@cite#1{[#1]} 
  \makeatother    
\begin{document}

  \renewcommand{\thefootnote}{\alph{footnote}}
\def\fsl#1{\setbox0=\hbox{$#1$}                 
   \dimen0=\wd0                                 
   \setbox1=\hbox{/} \dimen1=\wd1               
   \ifdim\dimen0>\dimen1                        
      \rlap{\hbox to \dimen0{\hfil/\hfil}}      
      #1                                        
   \else                                        
      \rlap{\hbox to \dimen1{\hfil$#1$\hfil}}   
      /                                         
   \fi}                                         %
\newcommand{\tr}{\mbox{tr}}
\newcommand{\VEV}[1]{\langle #1 \rangle}
\newcommand{\gtrsim}{\mathop{>}\limits_{\displaystyle{\sim}}}
\newcommand{\lessim}{\mathop{<}\limits_{\displaystyle{\sim}}}
\newcommand{\UV}{\Lambda_{\rm UV}}
\newcommand{\IR}{\Lambda_{\rm IR}}
\newcommand{\NDA}{\Omega_{\rm NDA}}
\newcommand{\Lg}{\Lambda_{Li}}
\newcommand{\Ly}{\Lambda_{LY}}
\newcommand{\Lq}{\Lambda_{L3}}
\newcommand{\DxSB}{D$\chi$SB}
\newcommand{\NKKg}{N_{\rm KK}^g}
\newcommand{\NKKb}{N_{\rm KK}^{gs}}
\newcommand{\NKKf}{N_{\rm KK}^f}
\newcommand{\NKKs}{N_{\rm KK}^h}
\newcommand{\NKKi}{N_{\rm KK}^i}
\newcommand{\tMAC}{\Lambda_{\rm tM}}

\title{Top mode standard model and extra dimensions\footnote{
Talk given by M.H. at {\it The 12th International Conference on
Supersymmetry and Unification of Fundamental Interactions (SUSY 2004), 
June 17-23, 2004, Tsukuba, Japan}}}

\author{MICHIO HASHIMOTO\footnote{
The present address is {\it Department of Applied Mathematics,
Western Science Centre, The University of Western Ontario,
London, ON, Canada, N6A 5B7.}}\footnote{The new e-mail address: mhashimo@uwo.ca}}
\address{Department of Physics, Pusan National University, 
         Pusan 609-735, Korea \\
{\tt E-mail: michioh@charm.phys.pusan.ac.jp}}

\author{MASAHARU TANABASHI}
\address{Department of Physics, Tohoku University, Sendai 980-8578, Japan\\
{\tt E-mail: tanabash@tuhep.phys.tohoku.ac.jp}}

\author{KOICHI YAMAWAKI}
\address{Department of Physics, Nagoya University, Nagoya 464-8602, Japan\\
{\tt E-mail: yamawaki@eken.phys.nagoya-u.ac.jp}}

\abstract{
We perform the most attractive channel (MAC) analysis in 
the top mode standard model with TeV-scale extra dimensions,
where the standard model gauge bosons and the third generation of
quarks and leptons are put in $D(=6,8,\cdots)$ dimensions.
In order to make the scenario viable,
only the attractive force of the top condensate should exceed 
the critical coupling, while other channels such as the bottom and
tau condensates should not.
It turns out that the top condensate can be the MAC for $D=8$,
whereas the tau condensation is favored for $D=6$.
On the basis of the renormalization group equations 
for the top Yukawa and Higgs quartic couplings,
we predict masses of the top quark and the Higgs boson 
for $D=8$ as $m_t=172-175$ GeV and $m_H=176-188$ GeV, respectively.
}

\normalsize\baselineskip=15pt

\section{Introduction}

What is the physics behind the electroweak symmetry breaking (EWSB)?
Why are the masses of W, Z, and the top quark exceptionally large
compared with those of other particles in the standard model (SM)?
In the framework of the top quark 
condensate~\cite{MTY89,Nambu89,Bardeen:1989ds}, 
which is often called the ``Top Mode Standard Model'' (TMSM), 
the chiral condensation of the top quark $\VEV{\bar{t}_L t_R} \ne 0$
triggers the EWSB and then the top acquires its large mass of 
the order of the EWSB scale.
The Higgs boson emerges as the scalar bound state of $\bar{t}t$.

Along with the TeV-scale extra dimension 
scenario~\cite{Antoniadis:1990ew,Dienes:1998vh},
the TMSM has been reconsidered by several authors.~\cite{Dobrescu:1998dg,Cheng:1999bg,Arkani-Hamed:2000hv,Hashimoto:2000uk,Gusynin:2002cu,Hashimoto:2003ve,Gusynin:2004jp}
In particular, Arkani-Hamed, Cheng, Dobrescu and Hall (ACDH) proposed a model 
where the SM gauge bosons and the third generation quarks/leptons
live in the $D(=6,8,\cdots)$-dimensional bulk.~\cite{Arkani-Hamed:2000hv}  
The full bulk gauge dynamics was analyzed in 
Refs.~\cite{Hashimoto:2000uk,Gusynin:2002cu}, 
based on the ladder Schwinger-Dyson (SD) equation.

We here study the phenomenological implications.
(For details see Ref.~\cite{Hashimoto:2003ve}. )
In order for only the top quark to acquire the dynamical mass of 
the order of the EWSB scale, 
the binding strength only for the top quark should exceed 
the critical one $\kappa_D^{\rm crit}$. 
We thus analyze binding strengths of top, bottom and tau condensates, 
$\kappa_{t,b,\tau}$,
by using renormalization group equations (RGEs) of 
bulk gauge couplings.
The RG flows are compared with the lowest possible values of
$\kappa_D^{\rm crit}$~\cite{Hashimoto:2000uk,Gusynin:2002cu}.
We then find that the simplest scenario with $D=6$ cannot work,
while the model with $D=8$ can do.
On the basis of RGEs, 
we predict the top quark mass $m_t$ and the Higgs boson mass $m_H$ 
for $D=8$ as $m_t=172-175$ GeV and $m_H=176-188$ GeV, respectively.

\section{tMAC analysis}

\begin{figure}[tbp]
  \begin{center}
  \resizebox{0.47\textwidth}{!}
            {\includegraphics{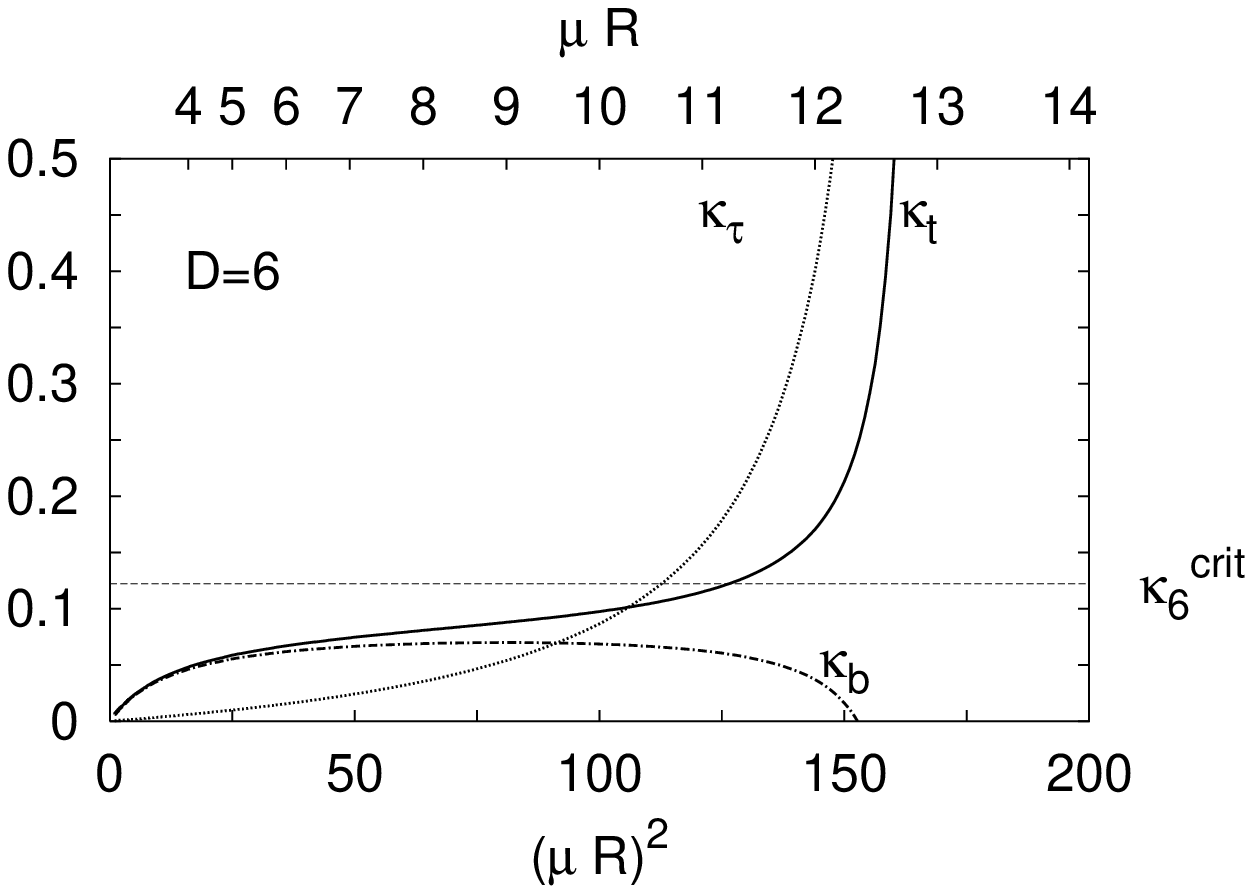}}\qquad
  \resizebox{0.47\textwidth}{!}
            {\includegraphics{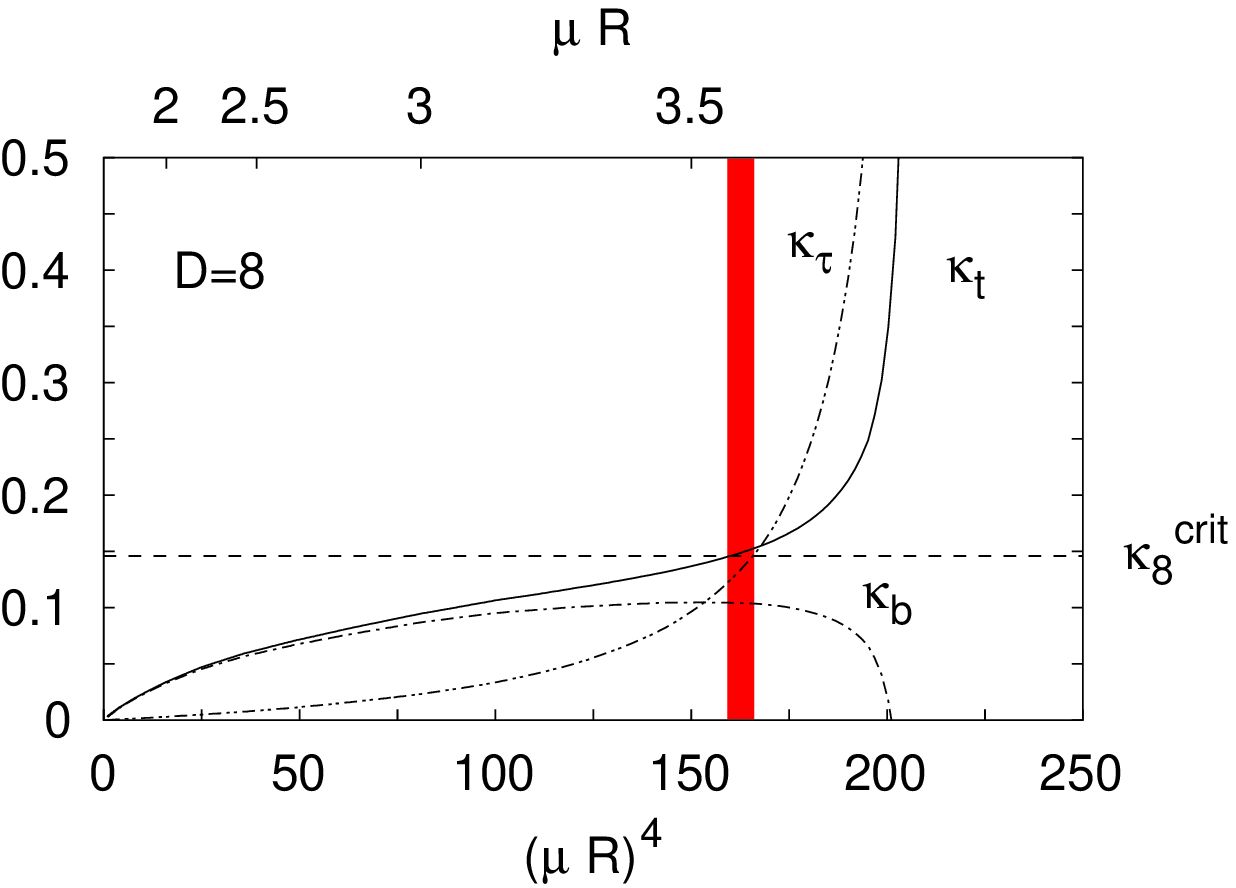}}
  \caption{Binding strengths $\kappa_{t,b,\tau}$ with 
           $R^{-1}=10$ TeV for $D=6,8$. 
           The critical binding strengths 
           $\kappa_{6,8}^{\rm crit}$ are shown
           by the horizontal dashed lines.
           The shaded region is the tMAC scale $\tMAC$ satisfying 
           Eq.~(\ref{top-cond-sec3}).
           \label{fig1}}
  \end{center}
\end{figure}

Let us consider a simple version of the TMSM with extra dimensions
where the SM gauge group and the third generation
of quarks and leptons are put in the bulk ($D=6,8,\cdots$), 
while the first and second generations live on the 3-brane 
(four dimensional Minkowski space-time). 
The extra $\delta (= D-4)$ spatial dimensions are compactified 
at a TeV-scale $R^{-1}$.
In order to obtain a four dimensional chiral theory and to forbid 
massless gauge scalars, we compactify extra dimensions on the orbifold 
$T^\delta/Z_2^{\delta/2}$.~\cite{Hashimoto:2000uk,Hashimoto:2003ve}

By using the ``truncated Kaluza-Klein (KK)'' 
effective theory~\cite{Dienes:1998vh},
we calculate the RGEs for the four dimensional gauge couplings 
$g_i (i=3,2,Y)$, (for details see Ref.~\cite{Hashimoto:2003ve})
\begin{equation}
  (4\pi)^2 \mu \frac{d g_i}{d \mu} = b_i\,g_i^3 
   + b_i^{\rm KK}(\mu)\,g_i^3, 
  \quad (\mu \geq R^{-1}) \label{rge_ED}
\end{equation}
with $b_3=-7, b_2=-\frac{19}{6}$ and $b_Y=\frac{41}{6}$,
where the RGE coefficients $b_i^{\rm KK}(\mu)$ 
are given by
\begin{align}
  b_3^{\rm KK}(\mu) &= -11\, \NKKg (\mu)
 +\frac{\delta}{2}\, \NKKb (\mu) 
 +\frac{8}{3}\, \NKKf (\mu), & \mbox{for} &\quad SU(3)_c, 
\label{b3KK} \\
  b_2^{\rm KK}(\mu) &= -\frac{22}{3} \, \NKKg (\mu)
 +\frac{\delta}{3} \, \NKKb (\mu) 
 +\frac{8}{3} \, \NKKf (\mu) 
 +\frac{1}{6} \, \NKKs (\mu), & \mbox{for} &\quad SU(2)_W, 
 \label{b2KK} \\
  b_Y^{\rm KK}(\mu) &= 
 \frac{40}{9} \, \NKKf (\mu)
 + \frac{1}{6} \, \NKKs (\mu), & \mbox{for} &\quad U(1)_Y. 
 \label{byKK}
\end{align}
$N_{\rm KK}^k (\mu), \; k=g,gs,f,h$ denote
the total numbers of KK modes below $\mu$ 
for gauge bosons, gauge scalars, 
Dirac (4-component) fermions, and composite Higgs bosons, 
respectively. 

We define the {\it dimensionless} bulk gauge couplings $\hat g$ 
as~\cite{Hashimoto:2000uk,Gusynin:2002cu,Hashimoto:2003ve}
\begin{equation}
  \hat g_i^2(\mu) = \frac{(2\pi R \mu)^\delta}{2^{\delta/2}}g_i^2 (\mu),
  \quad (i=3,2,Y).
  \label{hat-g}
\end{equation}
Combining Eq.~(\ref{hat-g}) with Eq.~(\ref{rge_ED}), 
we find the RGEs for $\hat g_i$,
\begin{equation}
 \mu \frac{d}{d \mu} \hat g_i = \frac{\delta}{2}\,\hat g_i
 + \frac{\hat g_i^3}{(4\pi)^2}\frac{2^{\delta/2}}{(2\pi R \mu)^\delta}
   \left[\,b_i + b_i^{\rm KK}(\mu) \,\right] . \label{rge_ED2}
\end{equation}

We now analyze the energy scale $\tMAC$ 
where the top condensate is the most attractive channel (MAC) 
and only in the $\bar{t}t$-channel 
the binding strength exceeds the critical value $\kappa_D^{\rm crit}$, 
(``topped MAC'' or ``tMAC'' scale), 
\begin{equation}
  \kappa_t(\tMAC) > \kappa_D^{\rm crit} > \kappa_b(\tMAC),
  \kappa_\tau(\tMAC), \cdots. \label{top-cond-sec3}
\end{equation}
The binding strengths $\kappa_{t,b,\tau}$ are given by
\begin{align}
  \kappa_t (\mu)&= \frac{4}{3}\,\hat g_3^2 (\mu) \NDA 
                 + \frac{1}{9}\,\hat g_Y^2 (\mu) \NDA \label{k_t},
  & \mbox{for} & \quad \bar{t}t, \\
  \kappa_b (\mu)&= \frac{4}{3}\,\hat g_3^2 (\mu) \NDA 
                 - \frac{1}{18}\,\hat g_Y^2 (\mu) \NDA \label{k_b},
  & \mbox{for} & \quad \bar{b}b, \\
  \kappa_\tau (\mu)&= \phantom{\frac{4}{3}\,\hat g_3^2 (\mu) \NDA + \;\;}
                   \frac{1}{2}\, \hat g_Y^2 (\mu) \NDA \label{k_tau},
  & \mbox{for} & \quad \bar{\tau}\tau. 
\end{align}
We estimate $\kappa_D^{\rm crit}$
through the ladder SD equation.~\cite{Hashimoto:2000uk,Gusynin:2002cu} 
The lowest possible values are
\begin{equation}
 \kappa_6^{\rm crit} \simeq 0.122, \qquad
 \kappa_8^{\rm crit} \simeq 0.146. \label{kd_sd1}
\end{equation}

We show the results of the tMAC analysis in Fig.~\ref{fig1}.
For $D=6$, we find that {\it the tMAC scale is squeezed out and 
the tau condensation is favored}. 
For $D=8$, on the other hand, it turns out that the tMAC scale $\tMAC$
satisfying Eq.~(\ref{top-cond-sec3}) does exist, 
\begin{equation}
 \tMAC R = 3.5\,\mbox{--}\,3.6, \quad \mbox{for} \quad 
 R^{-1}=\mbox{1--100 TeV}.
 \label{tmac} 
\end{equation}

\begin{figure}[tbp]
  \begin{center}
  \resizebox{0.47\textwidth}{!}
            {\includegraphics{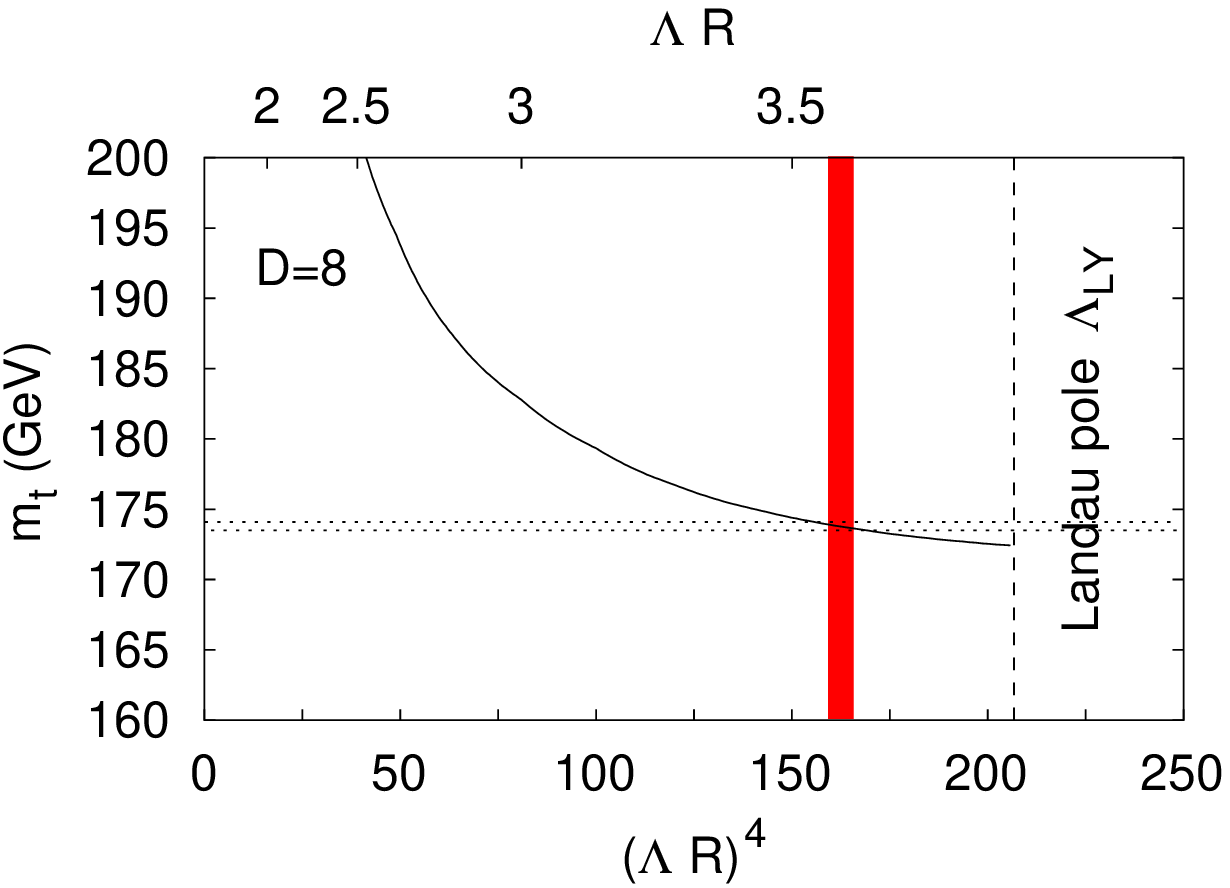}} \qquad
  \resizebox{0.47\textwidth}{!}
            {\includegraphics{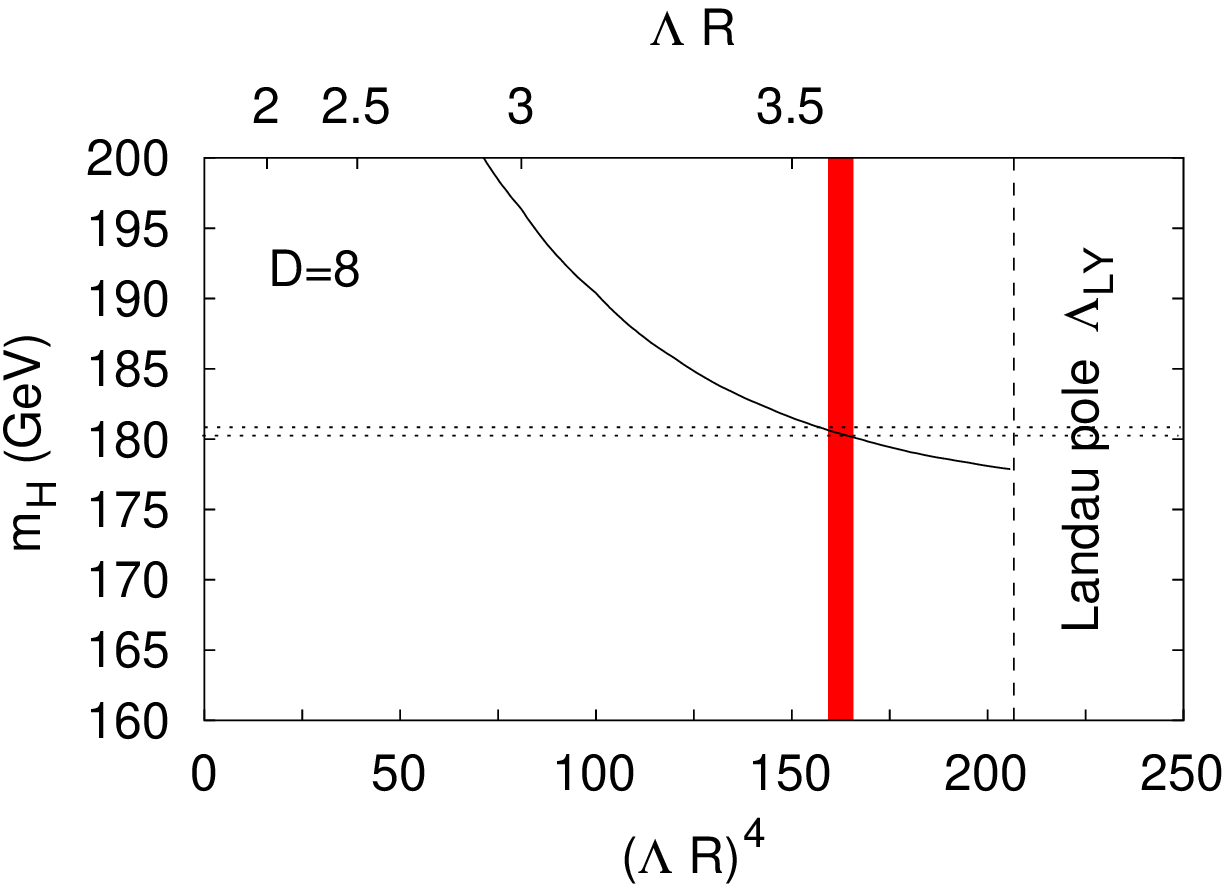}}
  \caption{Predictions of $m_t$ and $m_H$
           for $D=8,R^{-1}=10$ TeV.
           The dashed vertical line represents the Landau pole $\Ly$.
           The shaded region is the tMAC scale $\tMAC$ satisfying 
           Eq.~(\ref{top-cond-sec3}). 
           \label{mt-mh}}
  \end{center}
\end{figure}

\section{Prediction of $m_t$ and $m_H$}

We predict the top quark mass $m_t$ and the Higgs boson mass $m_H$ 
in a way used by ACDH~\cite{Arkani-Hamed:2000hv}.
This is similar to the approach of Bardeen, Hill and 
Lindner~\cite{Bardeen:1989ds}:
The masses $m_t$ and $m_H$ are predicted by using the RGEs of 
the top Yukawa and Higgs quartic couplings ($y$ and $\lambda$)
with compositeness conditions,
\begin{equation}
 y(\mu) \to \infty , 
 \quad \frac{\lambda(\mu)}{y(\mu)^4} \to 0 , \quad
 (\mu \to \Lambda). \label{comp-cond}
\end{equation}
While the compositeness scale $\Lambda$ 
in Ref.~\cite{Arkani-Hamed:2000hv} was treated 
as a free parameter to be adjusted for reproducing
the experimental value of $m_t$, 
we identify $\Lambda$ with the tMAC scale $\tMAC$
and hence $\Lambda$ is no longer an adjustable parameter 
but constrained as Eq.~(\ref{tmac}).
Thus we can test our model by comparing the predicted $m_t$ with 
the experimental value.

We show the results of $m_t$ and $m_H$ 
for $D=8$, $R^{-1}=10$ TeV in Fig.~\ref{mt-mh}, 
where the tMAC scale $\tMAC$ is shown by the shaded region. 
We then predict $m_t$ and $m_H$ as
\begin{equation}
  m_t = 172-175 \; \mbox{GeV}, \quad 
  m_H=176-188 \; \mbox{GeV}, \quad \mbox{for} \quad
  R^{-1}=\mbox{1--100 TeV}.
  \label{8D-mt-mh}
\end{equation}
The predicted value of $m_t$ is acceptable.
The prediction for the mass of the composite Higgs boson
shown in Eq.~(\ref{8D-mt-mh}) 
can be tested in collider experiments such as LHC.

\section{Summary and discussions}

We have performed the tMAC analysis.
We found that 
the region of the tMAC scale is squeezed out for $D=6$, 
while it does exist for $D=8$, $\tMAC =$ (3.5--3.6)$R^{-1}$.
The prediction of the top quark mass for $D=8$ is successful
and the (composite) Higgs boson with the characteristic mass shown in
Eq.~(\ref{8D-mt-mh}) should be discovered at LHC.

For a viable model with $D=6$ the gauged Nambu-Jona-Lasinio (GNJL) 
model in the bulk may be helpful.~\cite{Gusynin:2004jp}

\bibliographystyle{plain}

\end{document}